# Improving efficiency of hospitals and healthcare centres

## Primary Focus on

- Key Performance Indicators For Healthcare Centres
  - Dashboards for Hospital Management

**Project Mentor**: Mr. Miten Mehta
Co-founder and Advisor,
KloudData Labs Private Limited,
Pune, India

**Project Guide and Certified by**: Mr. Supratim Biswas
Professor, Computer Science and Engg. Department,
Indian Institute of Technology, Bombay (IIT B).
Mumbai, India.

**Project by:** **Prachi Kariya**
**Standard XII**
**PACE Jr. Science College**
**Mumbai, India**

# Improving efficiency of hospitals and healthcare centres

## INDEX

## Abstract

The Project aims at improving the efficiency of hospitals and healthcare centres using Big Data Analytics to evaluate identified KPIs (Key Performance Indicators) of its various functions. The Dashboards designed using computer technology serves as an interactive and dynamic tool for various stakeholders, which helps in optimising performance of various functions and more so maximise the financial returns. The Project entails improving performance of patient servicing, operations and OPD departments, finance function, procurement function, HR function, etc. I developed KPIs and drilldown KPIs for various functions and assisted in designing and developing interactive Dashboards with dynamic charts.

# Acknowledgements:

First and foremost, my sincere thanks to Prof. Supratim Biswas, Computer Science and Engineering Department, IIT Bombay, India for being my guide for this Project. I sincerely appreciate his invaluable guidance, direction, support and inputs in preparing this project report. Under his guidance and training my understanding of the subject and knowledge of analytics increased manifold. He has been an invaluable source of inspiration and encouragement. He has always been there whenever I needed his direction. I am honoured and delighted to have the opportunity of working with Mr. Biswas and I am grateful to him for all the support that he extended to me.

I am grateful to Mr. Miten Mehta, Co-founder and Advisor of KloudData Labs Private Limited, and Mr. Mayur Umbarkar, Manager, India Business Intelligence Practice. During my internship at KloudData, I was assisted and guided on technical aspects of health care data collation, data analytics and Dashboard designing for improving efficiency of hospitals and healthcare centres.

And finally, I thank my parents for being with me all the time, more importantly during late hours and standing beside me during this journey.

**Prachi  Kariya**
**August 2015**

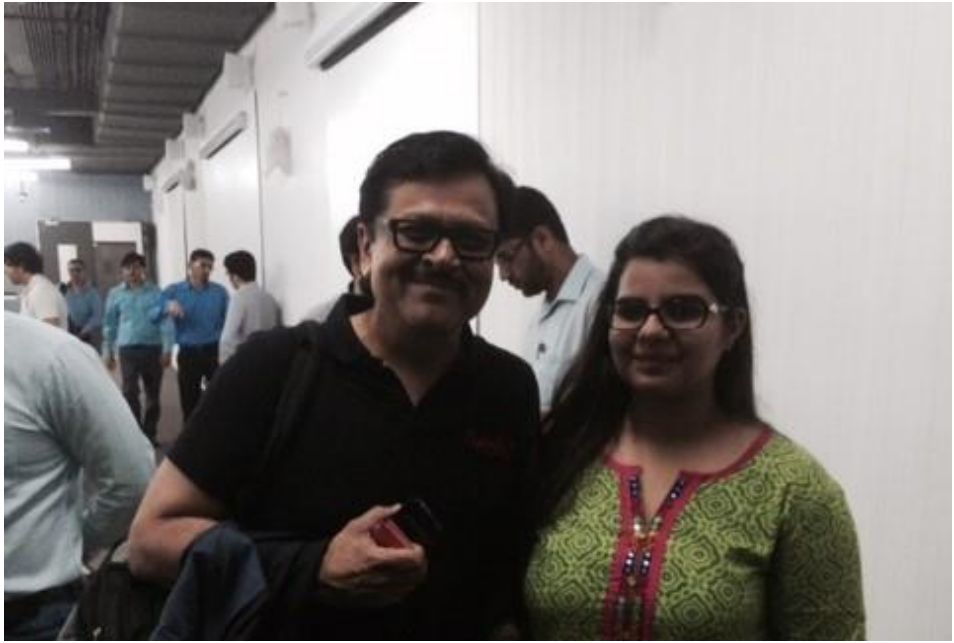

Prachi Kariya with Prashant Parekh, Founder and CEO, KloudData

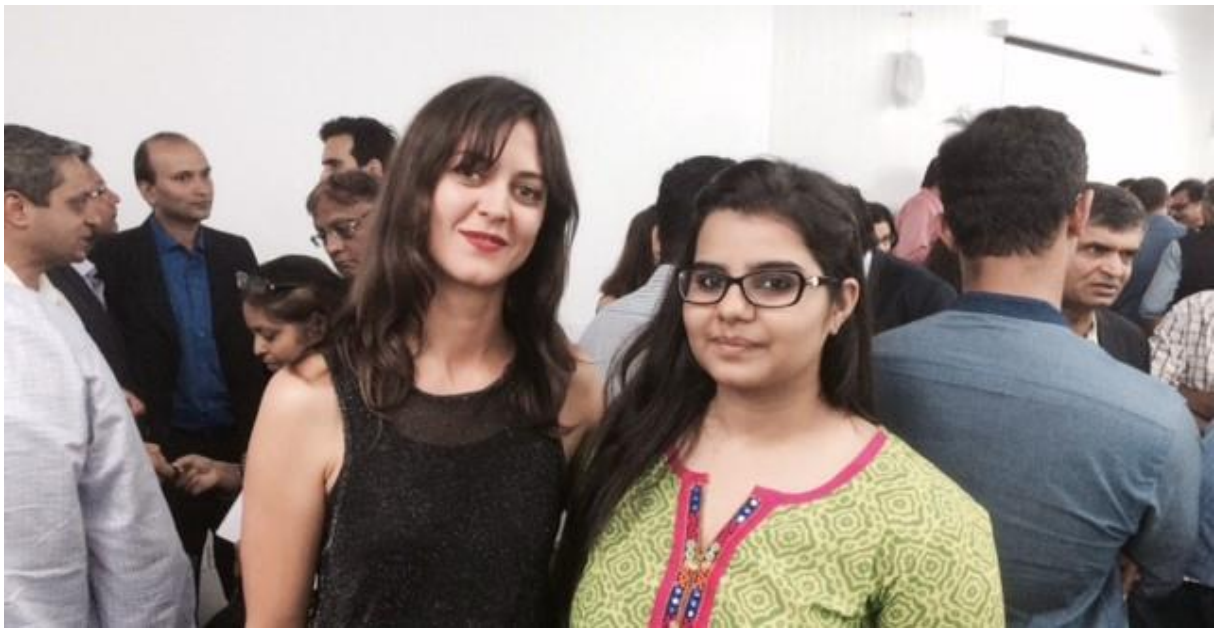

Prachi Kariya with Julia Watson, Founder of REDE Studios

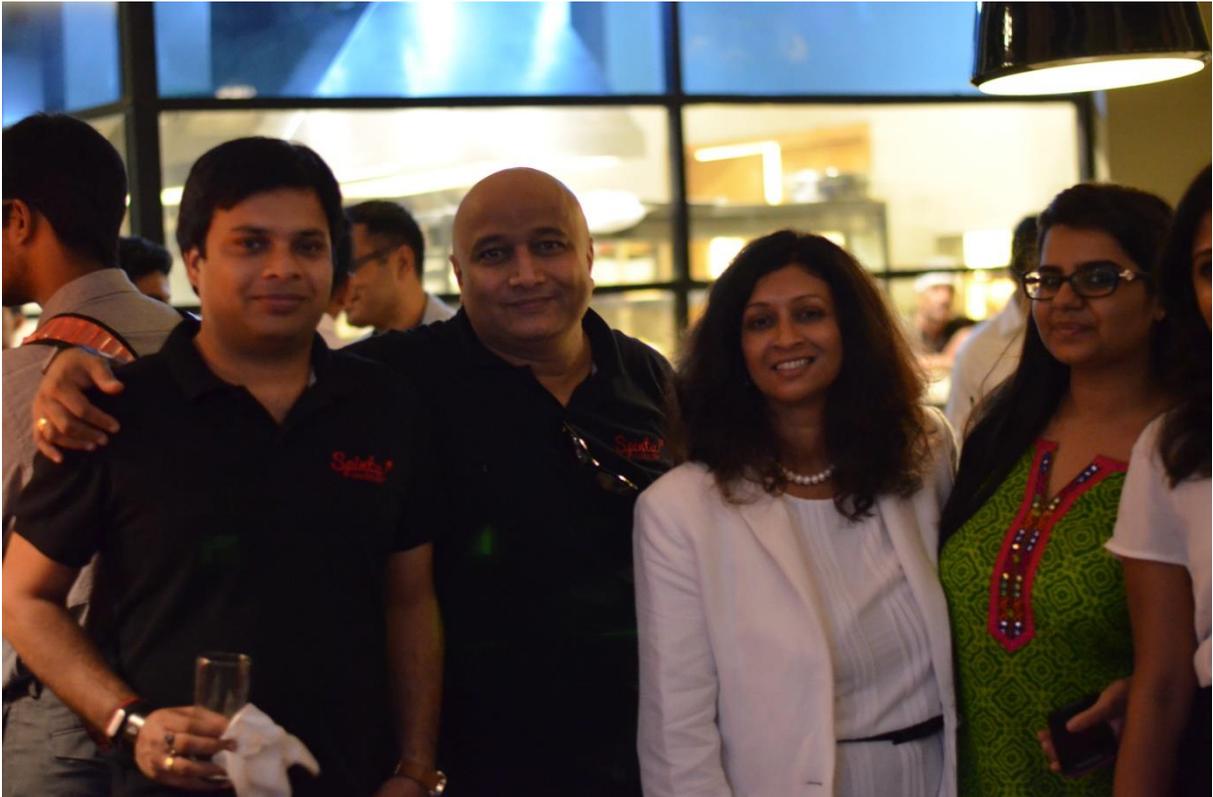

Prachi Kariya with Mr. Miten Mehta, Co-founder and Advisor, KloudData and the team of KloudData

# Acronyms

| Abbreviations | Full form |
| --- | --- |
| A/R | Accounts Receivables |
| CEO | Chief Executive Officer |
| CFO | Chief Financial Officer |
| CIT | Cold Ischemia Time |
| DRG | Diagnosis Related Group |
| EBITDA | Earnings before Interests, Taxes, Depreciation and Amortization |
| ER | Emergency Room |
| FTE | Full Time Equivalent |
| KPI | Key Performance Indicators |
| OR | Operations Room |
| PBT | Profit before Tax |
| POS | Point of Sale |
| RN | Resident Nurses |
| RVU | Relative Value Unit |
| ToT | Turnaround Time |
| YTD | Year to Date |

# Formulae

| Sr. | Parameter | Formula |
|-----|-----------|---------|
| 1. | EBITDA | Revenue – Expenses (Excluding tax, interest, depreciation and amortization) |
| 2. | EBITDA margin | EBITDA / Total Revenue |
| 3. | Operating Margin | Earnings Before Interest and Tax / Revenue |
| 4. | Earnings per share | Net Income / Number of shares |
| 5. | Return On Capital | Earnings Before Interest and Tax / Capital Employed |
| 6. | Return On Asset | Net Income / Average Total Assets |
| 7. | Days Cash On Hand | Cash / ( Operating Expense – Depreciation ) |
| 8. | Current Ratio | Current Asset / Current Liabilities |
| 9. | Debt Equity Ratio | Total Liabilities / Shareholders Equity |
| 10. | Collection Ratio | ( Debtors * Number of working days ) / Net Credit Sales |

## 1.     Healthcare Scenario

**1.1     Growth in Healthcare Sector:**

The modern and the present lifestyle has thrown various health related challenges for people across the globe and the citizens, the Government agencies and the healthcare organisations are very much concerned about the healthcare related issues and the proactive measures to be taken to improve the healthcare scenario. The health related problems differ dramatically from developed countries to the developing countries and similarly from urban centres to the underdeveloped rural areas.

The health problems suffered by people living in urban centres are more in the nature of high blood pressure, diabetes, heart attack etc. which are largely caused due to the increased stress level and hypertension, whereas, people in the underdeveloped rural areas have more of physical health related problems like skin disease, pneumonia, typhoid, etc. largely due to unhygienic living conditions and lack of healthcare facility.

Given this scenario, the healthcare sector has gained prominence around the world and most of the countries have started focussing on establishing healthcare infrastructure by taking various measures such as developing healthcare policies and programs, creating and allocating healthcare budgets and funds, establishing public healthcare facilities and healthcare organisations, encouraging private participation in developing healthcare facilities, incentivising and supporting innovating in various aspects related to healthcare, increasing the use of IT and telecom technology in effectively managing and providing high end healthcare facilities.

Globally, the healthcare sector is expanding at a rapid pace and the Government as well as private expenditure on the healthcare spend is increasing phenomenally as a percentage of Gross Domestic Product (GDP) of every country. The Economist Intelligence Unit (EIU) of World Health Outlook estimated that average global health care spending as a percentage of GDP will be 10.5 percent in 2014. The regional break-up estimates comprised of 17.4 percent for North America, 8.0 percent for Latin America, 10.7 percent for Western Europe, 6.6 percent for Asia/Australasia and 6.4 percent for the Middle East & Africa. (As per World Health Outlook 2013). As per the World Bank Report, the average global healthcare expenditure for the year 2013 was 9.8 per cent of GDP. The healthcare spend of developed countries like USA is as high as 17.1 per cent of GDP whereas that of a developing country like India and China is as low as 4 percent and 5.5 percent respectively of their GDP.

The healthcare spend between the private and the public sector depicts a skewed picture across geographies. Globally the average public sector spend on healthcare is many fold higher than that of private sector spending. The analysis reveals that the public sector spending in developed countries is on a higher side as compared to developing countries. As an example, the public sector spending in UK is as high as 7.6 per cent of its GDP as compared to India which is as low as 1.3 per cent.

**1.2    Market Scenario:**

Globally the healthcare expenditure is rising twice as fast as overall economic growth and at the same time the global healthcare industry is moving from a volume-based model to a value based business model. This requires the Government and the healthcare service providers to shift the gears towards hi-tech infrastructure enabled with sophisticated IT solutions and leveraging data analytics to improve the outcome and efficiency. Presently, the healthcare sector like other sectors are also suffering from the piles of fragmented data, that too housed at multiple sources and lacks quality. Therefore, Big Data and Data Analytics are not only gaining importance but has become a critical tool to improve efficiency and manage healthcare resources effectively. The global healthcare analytics market is expected to grow USD 21.3 billion by 2020.

**<<< This space is intentionally left blank >>>**

## 2.    Healthcare Goals

The Government as well as the private sector entrepreneurs operating healthcare facilities are focussed on effectively achieving following key goals to manage the healthcare infrastructure efficiently. Achievement of these goals is more critical for the public healthcare sector facilities operator given the higher % of public sector spending globally.

- ➢ Improve Operational Effectiveness
- ➢ Improve the quality of services in a time bound manner
- ➢ Reduce medication Errors
- ➢ Improve clinical effectiveness
- ➢ Use Business Intelligence and Data Analytics
- ➢ Improve financial and administrative performance
- ➢ Reduce readmissions
- ➢ Enhance member/ patient satisfaction

In order to achieve these goals, the management and hospital staff should perform various complex tasks by keeping pace with the dynamic healthcare environment including the regulatory changes. The complexity increases given the patient volumes and the types of patients, increasing supply costs, stringent Government compliances, quality requirements, multiple usage of assets and resources and finally the scarcity of trained staff.

The disjointed data located across various departments of a healthcare centre causes the biggest challenge in making the right decision at the right time. Non-availability of quality real time data and the data analytics does not give true picture of hospitals' performance with regard to operational, clinical and financial Key Performance Indicators (KPIs). Most hospitals have business intelligence systems that provide post facto analysis as opposed to a real time data analytics and the hospital executives have to depend on studying fragmented data across various departments which create bottlenecks in taking real time and effective decision making as well as problem solving.

In order to collate and analyse this data in a meaningful manner, it is important that the data structure should be KPI centric and more specific.

**<<< This space is intentionally left blank >>>**

## 3.    Key Performance Indicators

### 3.1    What are KPIs?

KPIs measure the performance and progress of an individual employee or the sector. KPIs are designed to measure the progress of the strategic goals decided and agreed as a part of the planning strategy. The use of KPIs has been in vogue for many decades where the same is being used by the government and public sector organisations over several years as part of the budgetary and reporting mechanism. The KPIs are an integral part of strategic and long term goal defining process. The effective KPI development should form a part of the planning and strategy, policy making, budgeting, review and monitoring.

Similar to goal setting, defining KPIs is also crucial as KPIs are in essence reflects a drilldown of specific goals in the form of activities that need to be performed to achieve the specific goals. Given this, the KPIs should also be SMART i.e.

S – Specific

M – Measurable

A – Achievable

R – Relevant

T – Time Bound

The KPIs are generally defined and agreed in the beginning of the year as part of the goal setting or budgetary planning. In many organisations, such KPIs are reviewed and redefined as part of the mid-term review process due to various factors such as –

- ➢ Change in the business strategy and policies of the organisation
- ➢ Change in the role of the staff
- ➢ Impact of the external factors
- ➢ Change in the regulatory framework
- ➢ Performance of the staff or the organisation
- ➢ Change in the goals of the staff

### 3.2    KPIs in Health Sector

Prof. Steve Rozner prepared a report titled "Developing Key Performance Indicators- A Toolkit for Health Sector Managers" with the support of USAID's Health Finance and Governance (HFG) and Abt Associates to improve the health sector in developing countries. The report is explained in detail about the KPIs, linking of KPIs to strategy, using KPIs in the healthcare sector, using health information systems to manage KPI data, etc.

In the said report, the author has explained that well defined KPIs should help the management team in doing number of things including:

➢ Establishing baseline information (i.e. the current state of performance);
➢ Setting up performance standards and targets to facilitate and encourage continuous improvement;
➢ Measuring and reporting improvements over a recurring intervals;
➢ Comparing performance across functions and geographic locations;
➢ Benchmarking performance against regional and international peers or norms;
➢ Allowing stakeholders to independently judge the performance of the healthcare centre or hospital.

The SMART KPIs help the Government including the Ministry of Health and Ministry of Finance to monitor the performance of the healthcare resources including the performance of the staff, improve the quality of service delivery, enhances overall efficiency by appropriately allocating and utilising the resources and improving the financial performance of the healthcare facilities.

**<<< This space is intentionally left blank >>>**

## 4.    Quality

### 4.1    Challenges:

Most of the hospitals face challenges on quality front and providing quality services is a very crucial aspect in the competitive environment. The issue of quality becomes graver when it comes to public hospitals. These aspects have significant bearing on the outcome of the services and satisfaction of the patients. Following are some of the quality challenges that need to be addressed on a priority basis.

- ➢ Drug safety and efficacy.
- ➢ Maintaining a safe, clean, hygienic and healthy environment.
- ➢ Quality of overall healthcare.
- ➢ Improved and efficient processes.
- ➢ Laboratory Accreditation.
- ➢ Continuous improvement of doctors, physicians, supporting staff.

### 4.2    Key Performance Indicators (KPIs)

| Sr. | KPIs | Drilldown KPIs |
|---|---|---|
| 1 | **HEALTHCARE MANAGEMENT** | |
| | **a)    External quality indicators** | ➢ Number of patients treated during the month and Year to Date (YTD) for different diseases such as pneumonia, stroke, cardiac surgery, nursing and its comparison with the set goals that reflects the quality level and the pressure on the resources of the hospital. |
| | **b)    Patient Survey** | ➢ Response of patients on various parameters that reflects on the quality of the services provided to the patient and the rating given by the patient on the overall performance of the hospital as well as hiss recommending the hospital to others. <br> ➢ Comparison of these parameters on month to month and YTD basis against the set goals |
| | **c)    Incidents** | ➢ Number of patients treated for various diseases. <br> ➢ Number of complaints received |
| | | and addressed across various departments on monthly and YTD basis on various parameters like professional conduct, patient communication and guidance, treatment and care, fast turnaround time (ToT), care wait time etc… |
| | | |
| 2 | **PATIENT SATISFACTION SURVEY ANALYSIS** | |
| | a) **Patient Care** | ➢ Feedback of patients on various aspects like consultation on arrival, diagnostics, nursing and treatment, medication etc… <br> ➢ Experience of the patient during his stay on parameters like personal attention, privacy, guidance and query handling, visiting time for friends and relatives, physiotherapy, diet etc… |
| | b) **Customer Service** | ➢ Feedback of the patient on the overall performance of the hospital on parameters like housekeeping, front office, billing, pharmacy etc… <br> ➢ Experience of the patient on softer aspects like courteousness of the staff, cleanliness during the stay, approach of support staff, provision of linen, friendly return policy for unused medicines upon discharge etc… |
| | | |
| 3 | **NURSING SCORECARD** | ➢ Feedback of people on Resident Nurse (RN) satisfaction scores, satisfaction and safety scores etc… |

# 5.    Operations

## 5.1    Challenges:

Efficient operation of hospitals is one of the major concern areas for the management team of any organisation as it involves multiple departments, numerous activities, intertwined processes, involvement of staff at different levels and more importantly efficient use of scarce resources. As a result of such diversity, the challenges are also diverse and of varied intensity, which requires detailing and working at various levels of the organisation. These challenges have a high impact on the productivity, employee morale and patient satisfaction. Following are some of the operational challenges that need to be addressed on a priority basis.

- ➢ Improve and optimise planning and scheduling.
- ➢ Maximisation of bed management and usage of hospital facilities.
- ➢ Understanding of Inpatient diagnosis and procedures with their cost.
- ➢ Streamlining and optimising utilisation of operation theatre.
- ➢ Streamlining and optimising utilisation of various assets including high-tech equipment.
- ➢ Wait time for patients at various departments and processes.
- ➢ Medication error.
- ➢ Average length of Stay vis-à-vis cost for the patient.
- ➢ Improved and efficient processes.
- ➢ Waste, fraud and abuse of resources.

## 5.2 Key Performance Indicators (KPIs)

| Sr. | KPIs | Drilldown KPIs |
|-----|------|----------------|
| 1 | **OPERATIONS MANAGEMENT** | |
| | **a)    Inpatient** | ➢ Quantitative parameters on number of admissions, unplanned re-admits, average length of stay, extended stay patients, long stay patients, wait time for admissions, emergency cases handled, etc. These numbers compared on monthly and YTD basis against the goal set.<br>➢ Capacity utilisation vis-à-vis target utilisation.<br>➢ Inpatient Diagnosis Related Groups (DRGs) including the number of cases attended and |
| | | revenues generated indicating most revenues generated, lowest margin and highest margin, etc. |
| **b)** | **Emergency Rooms (ER)** | ➤ Quantitative measurements around ER admit, ER presents, Diverts, time taken in ER, time to treatment, etc. These numbers compared on monthly and YTD basis against the goal set.<br>➤ ER Capacity utilisation vis-à-vis target utilisation.<br>➤ ER Procedure indicators like procedure of cases handled most revenues, lowest margin, highest margin, etc. |
| **c)** | **Surgery** | ➤ Quantitative measurements around total number of surgeries, OR waiting time and waiting list, OR utilisation and idle time as well as its comparison with the Goal set.<br>➤ Operation Theatre capacity utilisation vis-à-vis target utilisation.<br>➤ Operation Theatre pre-operative time and idle time<br>➤ Surgical procedure indicators like procedure of cases handled most revenues, lowest margin, highest margin, etc. |
| **d)** | **Outpatient** | ➤ Quantitative measurements around total number of outpatient admits, Relative Value Units (RVUs), average appointment wait time, registration wait time, no shows, operation time and lead time.<br>➤ Achievements across various indicators against the Goal set.<br>➤ Reduction in and control over number of "No Show"<br>➤ Capacity utilisation vis-à-vis target utilisation.<br>➤ Outpatient procedure indicators like procedure of cases handled most revenues, lowest margin, highest margin, etc. |
| **2** | **HOSPITAL PROCESS TIME ANALYSIS** | |
| | **a)    Clinical** | ➢ KPIs like initial assessment, patient information to the consultant, bed allocation time, first inward assessment, reporting investigation results, diagnostic analysis and treatment etc… <br> ➢ Achieving good results against ER and Non ER parameters vis-à-vis target set. <br> ➢ Achieving minimum time lag between various activities and processes mentioned above. |
| | **b)    Non clinical** | ➢ KPIs like pre-authorisation processes, counselling of relatives (timing, sensitivity and quality), quality of food, beverages and miscellaneous services, timely medication, discharge process including billing and payment <br> ➢ Achieving good results against ER and Non ER parameters vis-à-vis target set. <br> ➢ Achieving minimum time lag between various activities and processes mentioned above. |
| | | |
| **3** | **WAITING LIST AND TRANSPLANT** | |
| | **a)    Cold Ischemia Time (CIT) Analysis** | ➢ Reducing cold ischemia time during the transplant operation to less than 9 hours to achieve a higher per cent of success rate. <br> ➢ Reducing the risk of failure by curtailing cold ischemia time during the operation. |
| | **b)    Transplant Analysis** | ➢ Reducing the overall waiting time for active and new additions. <br> ➢ Reduce transplant failure. <br> ➢ Successfully achieve the transplant using the organs of living donors. |

## Why-Because analysis-MRD

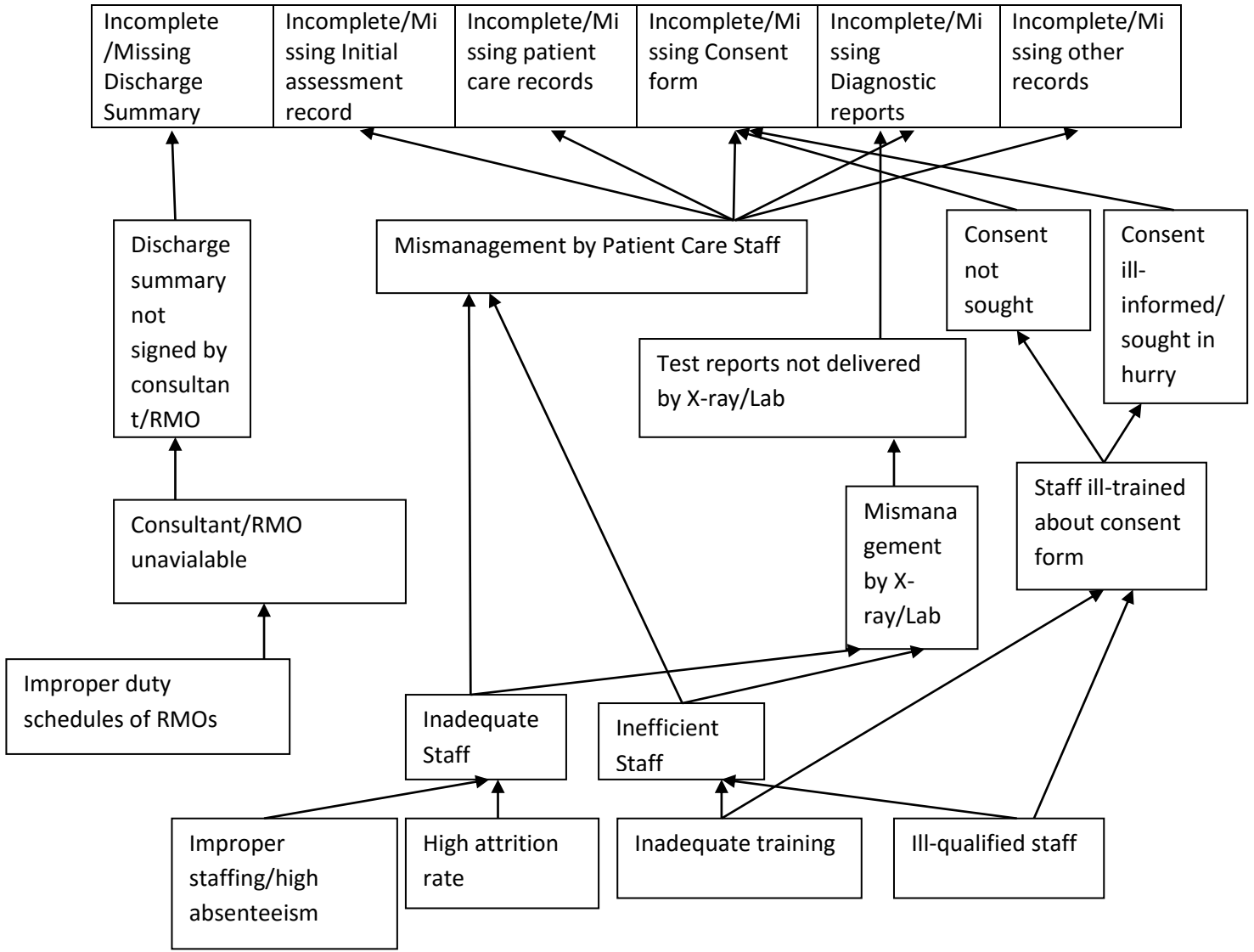

## 6.    Finance

### 6.1    Challenges:

Managing financial aspects of a hospital is a crucial function for the Government as well as the private sector operators. Financially well managed hospitals not only help in providing cost effective healthcare solutions but also provides quality services to the patients at an affordable price. Developing and maintaining hospitals is a capital intensive affair and therefore managing costs, achieving profitability and sustainable growth are very important for any successful hospital venture. Apart from the profitability, maintaining cash flow and working capital for day to day smooth functioning of a hospital is a key to successful running of a hospital. These challenges have a significant impact on the working of the hospital as well as the credibility of the hospital. Following are some of the financial challenges that need to be addressed on a priority basis.

➢ Cash flow and working capital management.
➢ Achieving operating profitability on a sustainable basis.
➢ Optimising resources and processes and thereby reducing overall cost.
➢ Improving overall margin.
➢ Derive financial performance indicators for different departments or service lines.
➢  Managing employee cost without affecting the staff turnover ratio.
➢ Analyse and optimise workforce and benefits.
➢ Managing receivables below 90 days.
➢ Improving and maintaining credit rating of the hospital by ensuring healthy financial ratios.
➢ Generating and retaining funds for future capitalisation, modernisation and expansion.
➢ Cash embezzlement, wastage, fraud and leakages.

### 6.2 Key Performance Indicators (KPIs)

| Sr. | KPI | Drilldown KPIs |
|---|---|---|
| 1 | **FINANCIAL MANAGEMENT** | |
| | **a)    Financial core measures** | ➢ Consistently achieving financial parameters on month on month basis against the budgeted one- EBITDA (Earnings before Interests, Taxes, Depreciation and Amortization) Margin, Operating Margin, Profit Before Tax (PBT), Earnings Per Share, Return On Capital, Return On Asset etc… |
| | | ➢ Increasing the revenue from key resources like Revenue per doctor, Revenue per physician, Revenue per bed, Revenue per operation, Revenue per patient, Revenue per Full Time Equivalent (FTE) etc…<br>➢ Reducing operating costs – both FTE cost and administrative cost, overtime cost etc…<br>➢ Reducing referral commission, reimbursements for physicians, cost of idle hours and time gap between processes. |
| | **b) Cash flow measures** | ➢ Improving the cash flow and working capital.<br>➢ Improving financial parameters like Days Cash On Hand, Current Ratio, Debt Equity Ratio and Collection Ratio. |
| | | |
| **2** | **REVENUE CYCLE METRICS** | |
| | **a) POS Collection** | ➢ Improving Point of Sale (POS) collection on month on month basis across locations.<br>➢ Improving quarter on quarter collection performances. |
| | **b) Account Receivables (A/R) Management** | ➢ Achieving collection target on sustainable basis and reducing the collection time below 90 days.<br>➢ Improving billing cycle and dispatch of billing.<br>➢ Reducing write off.<br>➢ Improving the cycle of days to bill statements and days to bill claims. |
| | **c) Denials** | ➢ Significantly reducing claim denials by insurance companies.<br>➢ Reducing deductions of claim amounts by insurance companies on account of procedural delays and improper documentation. |
| | **d) Cash Receipts** | ➢ Minimising cash receipts at the registration and cash counters.<br>➢ Depositing cash received from patients into the bank account on the same or next day. |
| | | ➢ Reducing cash reimbursements to physicians and staff. |
| | | |
| 3 | **REVENUE AND OCCUPANCY BY DOCTORS/PHYSICIANS** | ➢ Increasing revenue performance per doctor by improving the processes and enhancing efficiency.<br>➢ Reducing cost per FTE of physicians and reimbursements.<br>➢ Increasing revenue margin per physician.<br>➢ Effectively managing the scheduling of doctors and physicians and increasing the revenue by carrying out more functions/operations by the doctors. |

**<<< This space is intentionally left blank >>>**

# 7.  Platform and Components

## 7.1    Challenges:

Managing financial aspects of a hospital is a crucial function for the Government as well as the private sector operators.

## 7.2    Components of Dashboard:

The key components of dashboard comprises of BI system, information data base of various departments, KPI database, IT platform and dynamic software. The pictorial representation of a sample dashboard is depicted below:

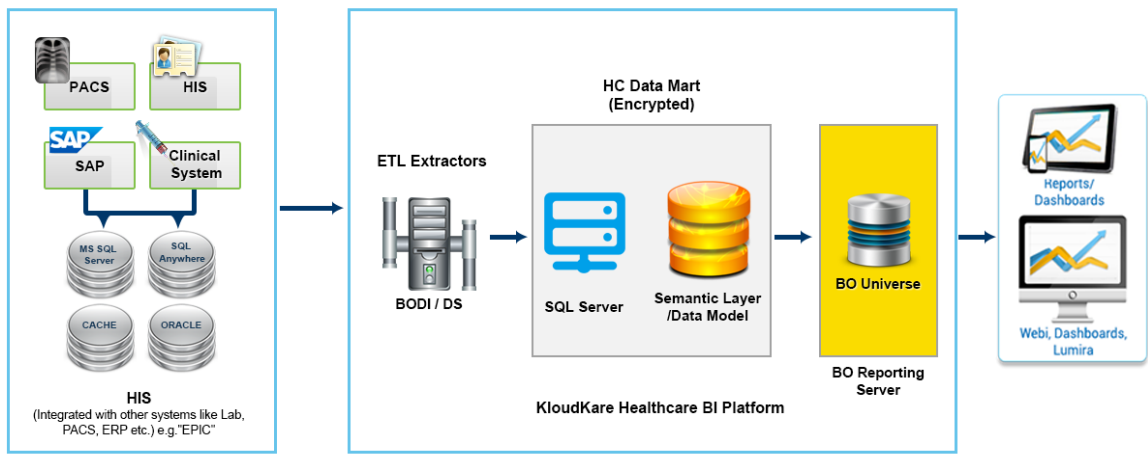

The following are the applications used for creating the Dashboards :

>    SAP Proprietary/Licensed Tools are used here.
> ➢ SAP Xcelcius Dashboard (Front End Client)
> ➢ SAP Information Designer Tool (IDT is middleware for Data Model)
> ➢ BOE Platform (Application Server)
> ➢ Database : MS SQL Server.
> ➢ Graphical User Interface : MATLAB

## 8.    Dashboards

Most of the hospitals have some or the other Business Intelligence System, which captures the data of carious functions and departments. This information is generally disjoint and disorganised so much that it does not give a comprehensive view of the operations of the hospital and does not help the management team in taking effective decisions. It does help in monitoring and improving the performance of a specific division but does not help to improve the overall efficiency.

In order to overcome these shortcomings, dashboards are being designed and developed using the existing BI system, data and information generated across various divisions, KPI linked performance indicator data, and the sophisticated IT platform. This dashboard provides a 360 Degree view of the hospital functioning to the COO and the CEO of the hospital that helps to identify and eliminate root causes of inefficiency and poor performance. The well designed and structured dashboard solution provides real-time information in a capsulated format that helps in accelerated and quality decision making process.

### 8.1    Features of effective Dashboard:

Development of an effective dashboard is as important as developing a markets strategy for the organisation. The well-structured and integrated dashboard provides en effective solutions to many of the problems of the healthcare sector. Some of the important characteristics of effective dashboards are highlighted below:

➢ Well defined KPIs for various functions and departments of hospitals;
➢ Capturing of quality and timely data and information from various BI systems used in the organisation
➢ Good BI toll components;
➢ Design should provide customised dashboards to Department Head of the specific functions like clinical function, operations function, finance function, human resource function. At the same time, the dashboard should also provide an overview of the functioning of entire hospitals to key management personnel like CEO and CFO.
➢ Provide timely alerts, escalations and indications
➢ Highlight trends and defined set of actions (manual or automated)
➢ Capable to access and interact with other medical systems
➢ Flexible, expandable and upgradable to meet future requirements
➢ Provide 360 Degree view and real time analysis of the functioning of the hospital
➢ Windows based design to facilitate navigating through the source database and KPI database

➢ It should be user friendly, provide summary and conclusions using visual aids, pictorial graphs, charts, etc.
➢ The dashboard should be interactive to facilitate quick decision making
➢ Data protection and strong security

The dashboard has been accepted as an effective tool to monitor operations of the hospital and has also helped in achieving the measurable savings in the forms of reduction in the cost of operation, improved efficiency, patient satisfaction, etc. The information rich dashboard enables and empowers the staff to take timely decisions and avoid eventualities and adverse events.

## 8.2    Sample reports generated from dashboard:

I have worked with KloudData team in the past in assisting them in conceptualising and developing the dashboard for healthcare sector. I was guided and supported by KloudData team that comprised of Mr. Miten Mehta, Co-founder and Advisor of KloudData Inc. and Mr. Mayur Umbarkar, Manager, India Business Intelligence Practice. Few sample reports generated from the dashboard are depicted below.

<u>**Executive Overview:**</u>

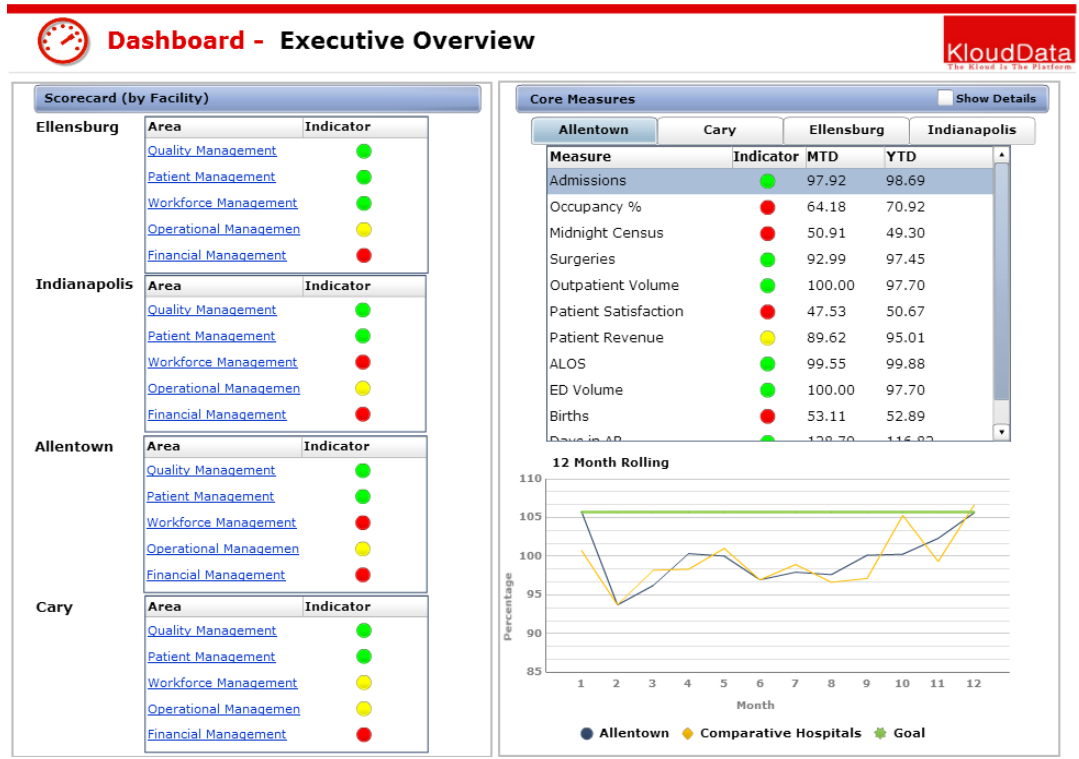

**Healthcare Management:**

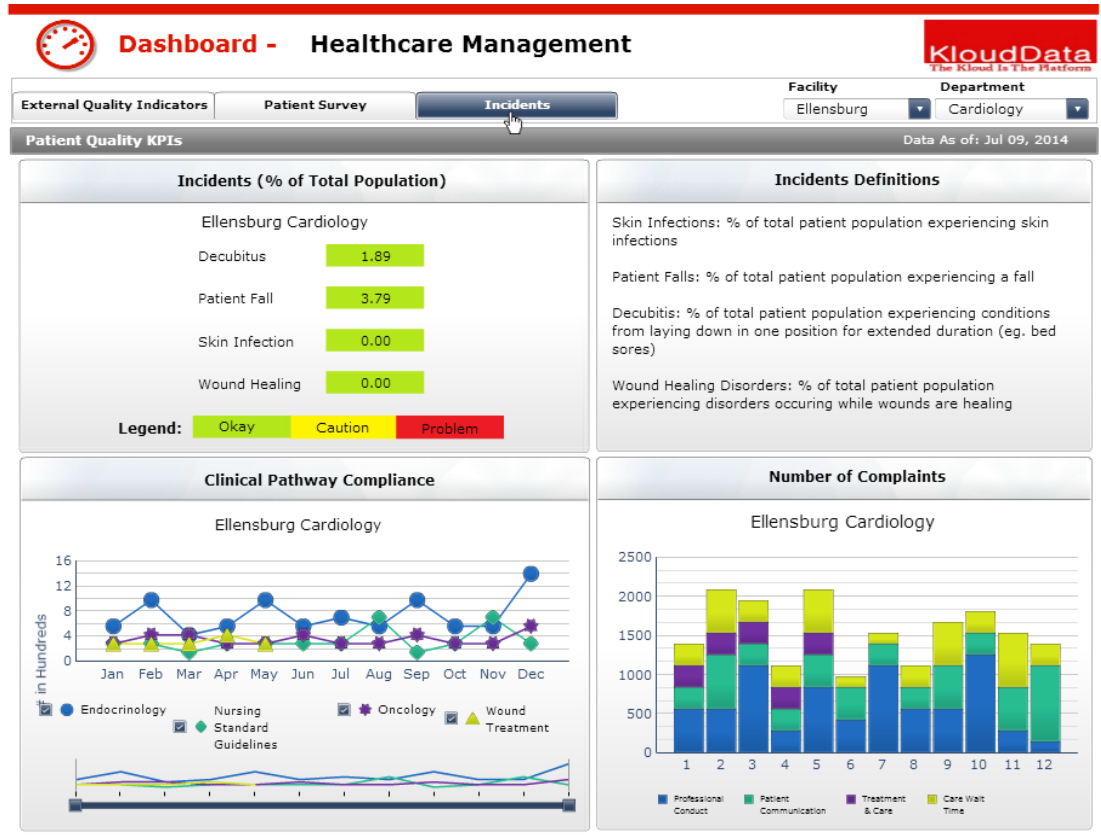

**Operational Management:**

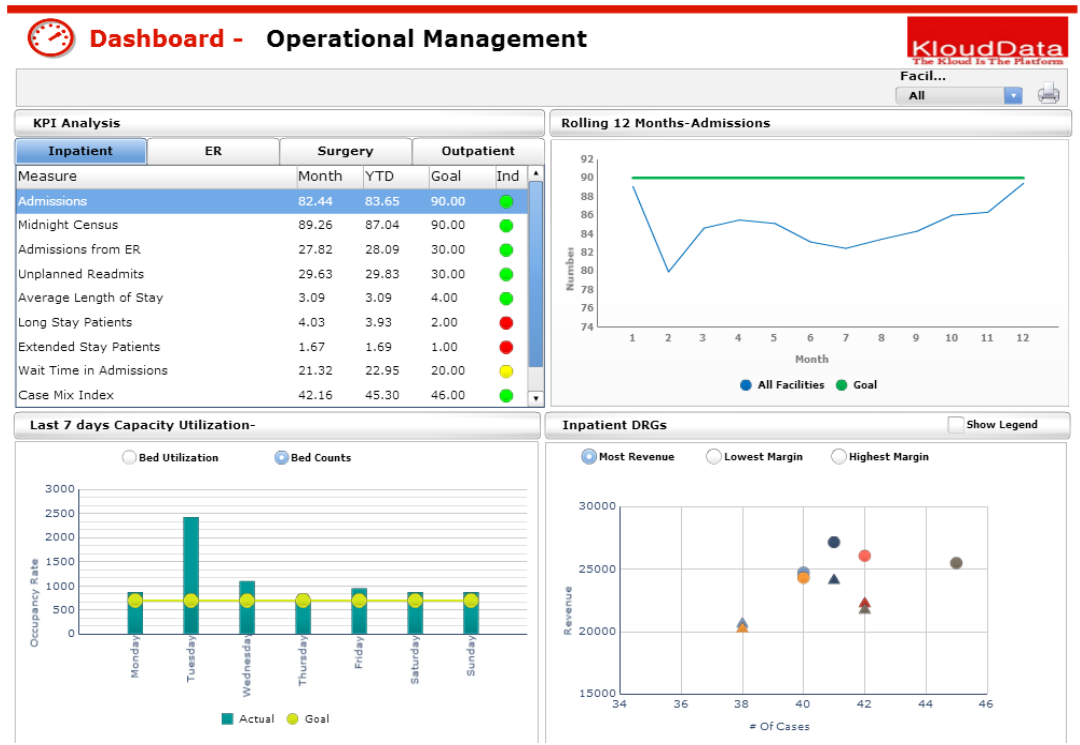

**Financial Management:**

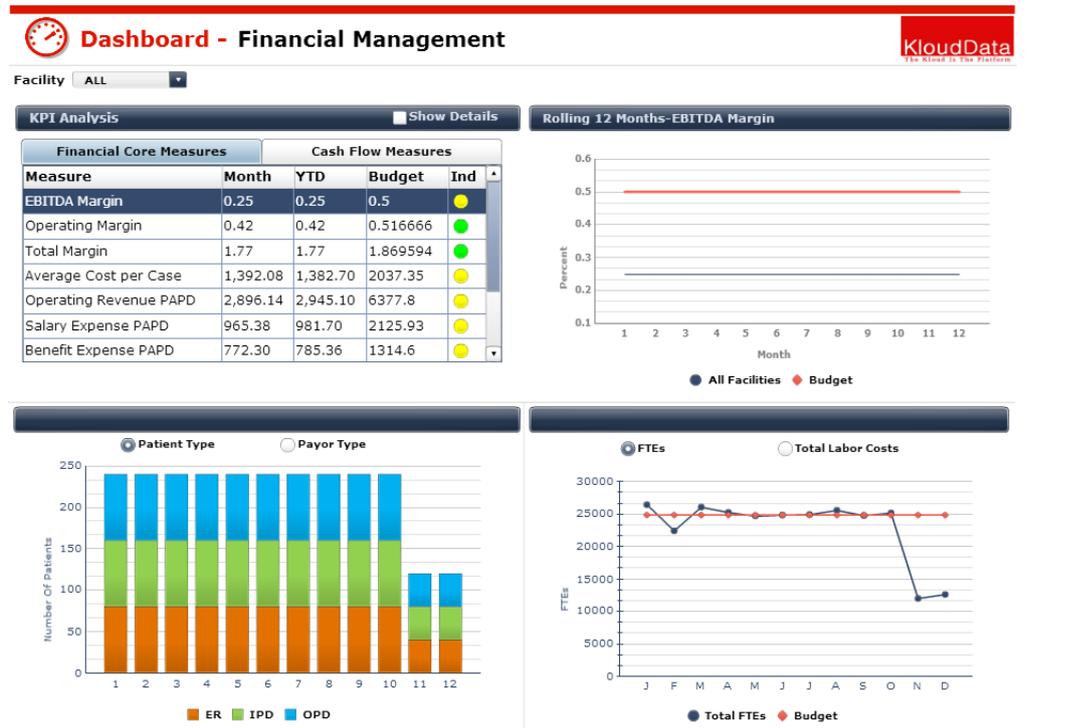

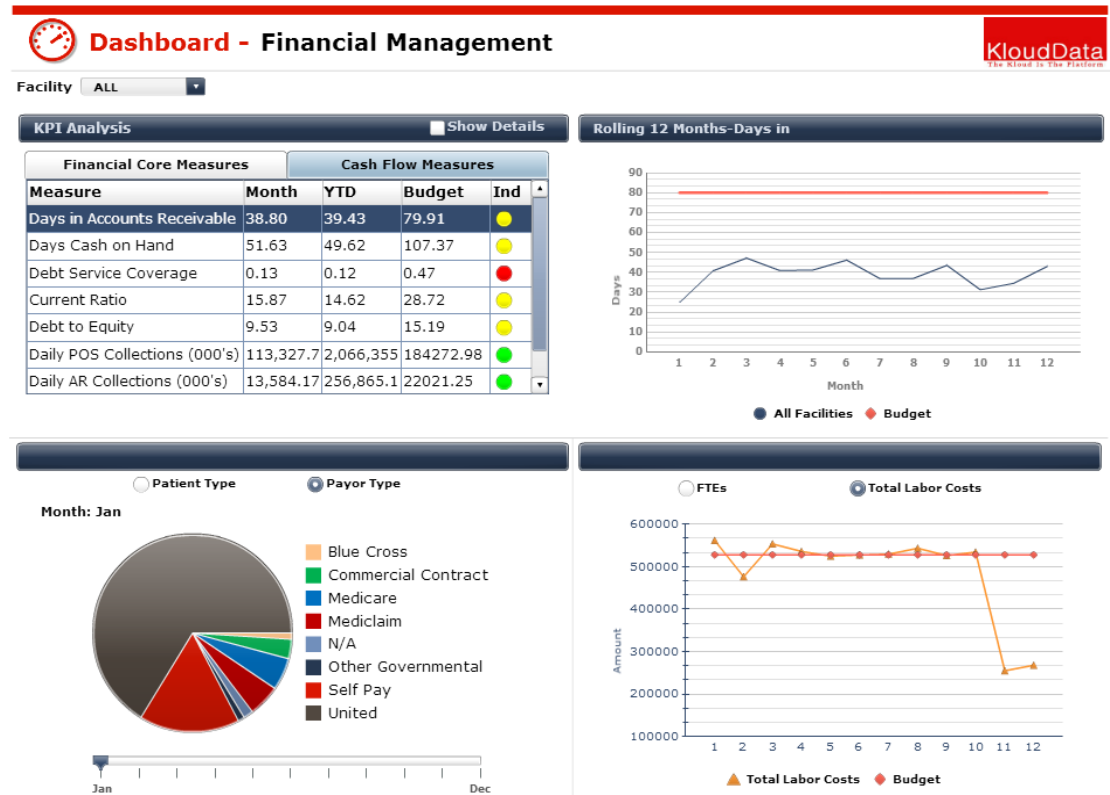